\documentclass[12pt,a4paper]{article}
\usepackage[utf8]{inputenc}
\usepackage[english]{babel}
\usepackage{amsmath}
\usepackage{cancel}
\usepackage{amsfonts}
\usepackage{amssymb}
\usepackage{graphicx}
\usepackage{float}
\usepackage{multirow}
\usepackage{wrapfig}
\usepackage{hyperref}

 \newtheorem{theorem}{Theorem}

\newcommand{\RR}{\mathbb{R}}

\newcommand{\norm}[1]{\left\Vert#1\right\Vert}
\newcommand{\abs}[1]{\left\vert#1\right\vert}

\title{Periodic solutions for the Lorentz force equation with singular potentials}

\author{Manuel Garz\'on and Pedro J. Torres\footnote{Partially supported by by MICENO and ERDF project MTM2017-82348-C2-1-P.} \\
 Departamento de Matem\'atica Aplicada, Universidad de Granada,\\
 Facultad de Ciencias, Granada, Spain.}

\begin{document}
\date{}
\maketitle

\begin{abstract}
We provide sufficient conditions for the existence of periodic solutions of the of the Lorentz force equation, which models the motion of a charged particle under the action of an electromagnetic fields. The basic assumptions cover relevant models with singularities like Coulomb-like electric potentials or the magnetic dipole. 
\end{abstract}



\section{Introduction}

The Lorentz force equation (LFE) models the motion of a slowly accelerated charged particle under the influence of an electromagnetic field. It is recognized as one of the fundamental equations of Mathematical Physics and the backbone of Electrodynamics \cite{F,J}. This system can be written as
\begin{eqnarray}\label{generalLFE}
\left(\phi(q')\right)'= E(t,q) + q'\times B(t,q).\end{eqnarray}
Here $E,B$ denote the electric and magnetic field respectively and $\phi$ is the relativistic acceleration operator given by 
\begin{eqnarray}
\phi(v)=\dfrac{v}{\sqrt{1-\abs{v}^2}},\nonumber
\end{eqnarray}
where $\abs{\cdot}$ stands for the usual euclidean norm in $\RR^3$. Without loss of genera\-li\-ty, the charge-to-mass ratio has been normalized to the unity. 

Due to its genuine theoretical and practical significance, the LFE has been studied from many perspectives. In particular, due to the known fact that a uniform constant magnetic field may induce a circular motion in a charged particle, it is natural to ask about sufficient conditions for the existence of closed trajectories. The existence of periodic solutions has been considered in recent works by means of topological \cite{BM2} or variational methods \cite{ABT,ABT2}. However, the cited references fail to cover the case of fields with singularities that are fundamental in Electromagnetism like Coulomb electric potential or the magnetic dipole. Our main objective is to fill, at least partially, this gap.

We consider an electric field of the form
\begin{eqnarray}
E(t,q)=-\nabla V(q) + h(t),\nonumber
\end{eqnarray}
for $V\in\mathcal{C}^1\left(\RR^3\setminus\{0\},\RR\right)$ and $h\in L^1\left(\left[0,T\right],\RR^3\right)$. Moreover, the potential $V$ satisfies the following assumptions:
\begin{itemize}
    \item[$(H1)$] $\lim_{|q|\rightarrow\infty}|\nabla V(q)|=0$. 
    \item[$(H2)$] $q\cdot\nabla V(q)$ is negative for any $q$ and there exist $c_0,\varepsilon_0>0$ and $\gamma\geq1$ such that $q\cdot\nabla V(q)\leq-c_0 |q|^{-\gamma}$ for any $|q|<\varepsilon_0$. 
\end{itemize}
The canonical model of an electric potential that verifies $(H1)-(H2)$ is the Coulomb potential $V(q)=c_0 |q|^{-1}$, i.e. the electric potential generated by a static charge placed at the origin.

On the other hand, the magnetic field $B\in\mathcal{C}\left([0,T]\times\mathbb{R}^3\setminus\{0\}\right)$ is required to verify the following hypotheses
\begin{itemize}
    \item [(H3)] There exists $C_B>0$ such that $\limsup_{|q|\rightarrow\infty} |B(t,q)|<C_B.$
    \item [(H4)] There exist $\varepsilon_1,c_1>0$ and $\beta\in(0,\gamma)$ such that $|B(t,q)|\leq c_1|q|^{-\beta-1}$ for all $t$ when $|q|<\varepsilon_1$.
\end{itemize}

Note that conditions $(H3)-(H4)$ are fulfilled in particular by a bounded magnetic force, like for instance an ABC magnetic field. In fields like Astrophysics and Plasma Physics, ABC magnetic fields play an important role and may serve as a relevant example of bounded magnetic fields. Even in the absence of electric field, it is known that ABC magnetic fields may generate complex dynamics, including chaotic motion and Arnold diffusion (see for instance \cite{D,LP1,LP2}). Other examples of bounded magnetic fields have been studied in the literature \cite{ANZZ,CP}, revealing a rich dynamics. Moreover, condition $(H4)$ enables a singularity of the magnetic field near the origin, and for instance the magnetic dipole field ( in its many variants \cite{SSK}) is covered by our assumptions.

By a $T$-periodic solution of system \eqref{generalLFE}, we mean a function $q:[0,T]\longrightarrow\mathbb{R}^3$ that verifies the system in the Carath\'eodory sense, such that $|q'(t)|<1$ for all $t$ and 
\begin{equation}\label{per}
q(0)=q(T),\qquad q'(0)=q'(T).
\end{equation}
From now on, the mean value of a given  function $h\in L^1\left(\left[0,T\right],\RR^3\right)$ is denoted by $\overline h=\frac{1}{T}\int_0^T h(t)dt$. Our main result is as follows.

\begin{theorem}\label{main-th1}.
Assume $(H1)-(H2)-(H3)-(H4)$. Then, for any $h\in L^1\left(\left[0,T\right],\RR^3\right)$ such that $\abs{\overline h}>C_B$, the Lorentz force equation \eqref{generalLFE} admits at least one $T$-periodic solution.
\end{theorem}

The proof relies on a global continuation theorem for periodic perturbations of autonomous system due to Capietto, Mawhin and Zanolin \cite{CMZ}. To this purpose, it is necessary to derive a priori bounds for the position and momentum of any eventual $T$-periodic solution of a suitable homotopic system that drives the original problem to an autonomous system. This is developed in Section \ref{ssec2} by means of similar techniques to that employed in \cite{HS,Zh}. In our case, the dissipative effect assumed in the cited references is replaced by the relativistic effect. The proof finishes in Section \ref{sec3} with the computation of the Brouwer degree of the vector field.

\section{A priori bounds}\label{ssec2}
We begin by defining the homotopic system:
\begin{equation}\label{Homotopy}
\left(\phi(q')\right)'= -\nabla V_\lambda(q) + h_\lambda(t) + \lambda \left(q'\times B(t,q)\right),\quad\lambda\in[0,1],
\end{equation}
where $V_\lambda(q)=\lambda V(q) + (1-\lambda)c_0|q|^{-1}$, $h_\lambda(t)= \lambda h(t) + (1-\lambda)\overline h$. It is clear that the original LFE corresponds to $\lambda=1$. 

The objective of this section is to find uniform (not depending on $\lambda$) a priori bounds for the $T$-periodic solutions of \eqref{Homotopy}.

\subsection{Upper bound}\label{up}
By $(H1)$ and $(H3)$, there exists $R>0$ (not depending on $\lambda$) such that $\abs{B(t,q)}<C_B$ and $\left|\nabla V_\lambda(q)\right|<|\overline h|-C_B$ for all $t\in[0,T]$ and any $\abs{q}>R$. Suppose that $q(t)$ is a solution of \eqref{Homotopy} such that doesn't belong to $B_R(0)$, for all $t$. Note that $B_R(0)$ denotes the ball of radius R and centered at the origin. Then, by integrating \eqref{Homotopy} in the whole period, we obtain
\begin{equation*}
    0=-\int_{0}^{T}\nabla V_\lambda(q(t))\ dt + \lambda\int_{0}^{T}q'(t)\times B(t,q(t))\ dt + T\overline h,
\end{equation*}
so
\begin{equation*}\left|\int_{0}^{T}\nabla V_\lambda(q(t)) dt\right|=
\left|T\overline h + \lambda\int_{0}^{T}q'(t)\times B(t,q(t))\ dt\right|.
\end{equation*}
Now, bounding both sides we get a contradiction:
\begin{eqnarray}
\left|\int_{0}^{T}\nabla V_\lambda(q) dt\right|&\leq&\int_{0}^{T}\left|\nabla V_\lambda(q) \right|dt<T\left(\abs{\overline h}-C_B\right), \nonumber\\
\left|T\overline h + \lambda\int_{0}^{T}q'\times B(t,q)\ dt\right|&\geq& T\abs{\overline h}-\lambda\left|\int_{0}^{T}q'\times B(t,q)\ dt\right|\geq T\left(\abs{\overline h}-C_B\right).\nonumber
\end{eqnarray}
Here, we have used that $\abs{q'(t)}<1$ for all $t$. Thus, there exists at least an instant $\tilde{t}$ such that $|q(\tilde{t})|\leq R$. By using again the bound for the derivative, we get
\begin{eqnarray}
|q(t)| = \left|q(\tilde{t}) + \int^{t}_{\tilde{t}}q'(s) ds\right|< R + T.\nonumber
\end{eqnarray}
Hence every $T$-periodic solution of \eqref{Homotopy} belongs to the open ball $B_{R+T}(0)$.

\subsection{Lower bound}\label{lower}
It's clear that, by conditions $(H2)$, exists a number $\varepsilon>0$ small enough such that
\begin{eqnarray}
-q\cdot\nabla V(q)\geq c_0|q|^{-1} + c_1|q|^{-\alpha},&
\text{for }\abs{q}<\varepsilon,\nonumber
\end{eqnarray} with $\gamma>\alpha>0$. Note that $\varepsilon$ depends of the constants $\alpha, c_0$ and $c_1.$ In particular, taking $\alpha=\beta$ we get:
\begin{eqnarray}
-q\cdot\nabla V_\lambda(q)+\lambda q\cdot v\times B(t,q)&\geq& c_0\left(\lambda\abs{q}^{-1}+(1-\lambda)\abs{q}^{-1}\right)= c_0\abs{q}^{-1},\ \ \ \ \label{ineq}
\end{eqnarray}
for $\abs q\leq\varepsilon$ and any $|v|\leq1$.

On the other hand, integrating the scalar product of \eqref{Homotopy} with a solution $q(t)$ over $[0,T]$, we obtain
\begin{eqnarray}
\int_{0}^{T}q(t)\cdot\left(\dfrac{q'(t)}{\sqrt{1-\left|q'(t)\right|^2}}\right)'dt&=&\int_{0}^{T}q(t)\cdot\left[h_\lambda(t) + \lambda q'(t)\times B(t,q(t))\right]dt \nonumber\\ &&-\int_{0}^{T}q(t)\cdot\nabla V_\lambda(q(t))dt.\nonumber
\end{eqnarray}
Integrating by parts on the left-hand side and using the periodicity of the solution, we see that it is a negative number, i.e:
\begin{eqnarray}
0&\geq& \int_0^T \left[-q(t)\cdot\nabla V_\lambda(q(t)) + \lambda q(t)\cdot q'(t)\times B(t,q(t))\right]dt\nonumber\\
&&+ \int_0^Tq(t)\cdot\left[\lambda h(t) + (1-\lambda)\overline{h}\right]dt.\nonumber
\end{eqnarray}
Thus, applying \eqref{ineq} we can write the following relation:
\begin{eqnarray}
I_\varepsilon(q(t))&=&\int_{\abs{q(t)}\leq\varepsilon}q(t)\cdot\left[-\nabla V_\lambda(q(t))+\lambda q'(t)\times B(t,q(t))\right]dt\nonumber\\
&\leq&\Bigg|\int_{\abs{q(t)}>\varepsilon}q(t)\cdot\left[-\nabla V_\lambda(q(t))+\lambda q'(t)\times B(t,q(t))\right]dt\nonumber\\
&&\ + \int_0^Tq(t)\cdot\left[\lambda h(t) + (1-\lambda)\overline{h}\right]dt\Bigg|\nonumber\\
&<&Tc_o\varepsilon
^{-1}+ TC_{\nabla V,B} + (R+T)\|h\|_1,\nonumber
\end{eqnarray}
where $$C_{\nabla V, B}:=\max_{\varepsilon<|q|<R+T}\left(\abs{\nabla V(q)} + \abs{B(t,q)}\right),$$
and $\|\cdot\|_1$ denotes the $L_1$ norm on $[0,T]$. This is a finite quantity because $\varepsilon$ is a fixed number and the fields $\nabla V$ and $B$ are continuous in that set. Futhermore, this estimation is independent of $\lambda$.

Now we assume that there is an interval $[t_1,t_2]$ where the particle enters the ball $B_{\varepsilon}(0)$, i.e:
\begin{eqnarray}
\big|q(t_1)\big|=\varepsilon,&\left|q(t)\right|<\varepsilon&,\forall t\in[t_1,t_2].\nonumber
\end{eqnarray}
Then, for all $t\in\left(t_1,t_2\right)$, we can write:
\begin{eqnarray}
\Big|\ln\left|q(t)\right|\Big|&=&\left|\ln\left|q(t_1)\right| + \int_{t_1}^{t}\dfrac{q(t)\cdot q'(t)}{\left|q(t)\right|^2}dt\right|\leq \Big|\ln|\varepsilon|\Big| + \int_{t_1}^{t}\left|q(t)\right|^{-1}dt\nonumber\\ 
&\leq& K_2+\int_{\abs{q(t)}\leq\varepsilon}\left|q(t)\right|^{-1}dt\leq K_2+\frac{1}{c_0}I_\varepsilon(q(t))\nonumber\\
&<& K_2 + T\varepsilon
^{-1}+ T\dfrac{C_{\nabla V,B}}{c_0}+\dfrac{R+T}{c_0}\|h\|_1.\nonumber
\end{eqnarray}
Because it is a finite bound, it follows that there exists a strictly positive lower estimate for the module of $q(t)$. More concretely,
$$
\abs{q(t)}>m:=\exp\left[-K_2- T\varepsilon
^{-1}- T\dfrac{C_{\nabla V,B}}{c_0}-\dfrac{R+T}{c_0}\|h\|_1 \right].
$$
So, we conclude that this constant $m$ bounds every periodic solution of \eqref{Homotopy} from below and, in particular, the singularity point is always avoided.

\subsection{Bound for the momentum}

For a given $T$-periodic solution $q(t)$ of the homotopic system \eqref{Homotopy}, the quantity $p(t)=\phi(q'(t))$ has a neat physical interpretation as the relativistic momentum of the particle. For our purposes, explicit bounds for the momentum will be needed as well. 

First, it is easy to verify that
\begin{eqnarray}
\abs{-\nabla V_\lambda (q)+\lambda \left(q'(t)\times B(t,q(t))\right)}\leq\abs{\nabla V(q)}+\frac{c_0}{\abs{q}^2}+ \left| B(t,q)\right|=:H(t,q).\label{momentum}
\end{eqnarray}
Then, by making use of the bounds obtained for the position of the particle $q(t)$, one has
$$
\abs{-\nabla V_\lambda (q)+\lambda \left(q'(t)\times B(t,q(t))\right)}\leq M=:\max_{\substack{m\leq\abs{q}\leq R+T,\\ t\in[0,T].}} H(t,q) .
$$
Assume now that $t_0\in [0,T]$ is a critical point of $q(t)$. Then,
\begin{eqnarray*}
\abs{p(t)} &=&\abs{\phi(q'(t))}=\abs{\int_{t_0}^t \left[-\nabla V_\lambda(q) + h_\lambda(s) + \lambda \left(q'\times B(s,q)\right)\right]ds}\\
&\leq&\int_0^T \abs{-\nabla V_\lambda(q) + h_\lambda(s) + \lambda \left(q'\times B(s,q)\right)}dt\\
&<& TM+2\norm{h}_1=: L.
\end{eqnarray*}
Of course, this estimate is independent of $\lambda$.

\section{Global continuation and topological degree}\label{sec3}

In this section, we are going to prove the main result by using a celebrated global continuation theorem due to Capietto, Mawhin and Zanolin \cite{CMZ}. We begin by writing system \eqref{Homotopy} as a first-order system with position $q$ and relativistic momentum $p=\phi(q')$ as the new coordinates, 
\begin{equation}\label{Hom-syst}
 \begin{split}
     q'  =  & \phi^{-1}(p),\\
      p'= & -\nabla V_\lambda(q) + h_\lambda(t) + \lambda \left( \phi^{-1}(p)\times B(t,q)\right).
   \end{split}
\end{equation}
This is a system of six differential equations. Let us define the Banach space $X=\left\{x\in C([0,T],\RR^6)\,:\, x(0)=x(T)\right\}$. By the results of Section \ref{ssec2}, every $T$-periodic solution of \eqref{Hom-syst} is contained 
into the open and bounded set
$$
\Omega=\left\{(p,q)\in X\,:\, m<\abs{q(t)}<R+T,\abs{p(t)}<1 \mbox{ for all }t\in [0,T]\right\}.
$$
Now, by defining $x=(p,q)$ and the function
$$
f(t,x;\lambda)=(\phi^{-1}(p),-\nabla V_\lambda(q) + h_\lambda(t) + \lambda \left( \phi^{-1}(p)\times B(t,q)\right),
$$
system \eqref{Hom-syst} reads simply
$$
x'=f(t,x,\lambda).
$$
To apply \cite[Theorem 2]{CMZ}, it remains to prove that
\begin{equation}\label{degree}
    d_B(f_0,\Omega\cap\RR^6,0)\ne 0,
\end{equation}
where 
$$
f_0(x)=f(t,x;0)=\left(\phi^{-1}(p),\overline h+c_0\frac{q}{\abs{q}^3}\right).
$$
Note that
$$\phi^{-1}(p)=\dfrac{p}{\sqrt{1+\abs{p}^2}},$$
hence $f_0$ is of class $C^\infty$ in the domain $\Omega\cap\mathbb{R}^6$. Then, we can calculate explicitly its partial derivatives:
\begin{eqnarray}
\dfrac{\partial f_0^i}{\partial p_j}\  (p,q)&=&\delta_{ij}\left(1+\abs{p}^2\right)^{-1/2} - p_ip_j\left(1+\abs{p}^2\right)^{-3/2},\nonumber\\
\dfrac{\partial f_0^{1+i}}{\partial q_j}(p,q)&=&\delta_{ij}\abs{q}^{-3}-3q_iq_j\abs{q}^{-5},\ \ \ \ \ \ i,j=1,2,3,\nonumber
\end{eqnarray}
where $\delta_{ij}$ denotes the Kronecker delta function. The other derivatives are identically zero, hence the Jacobian matrix is a block matrix such that
\begin{eqnarray}
\det \text{Jac} f_0 (p,q) =-2\abs{q}^{-9}\left[\left(1+\abs{p}^2\right)^{-3/2}-\abs{p}^2\left(1+\abs{p}^2\right)^{-5/2}\right].\label{Jacobian}
\end{eqnarray}
On the other hand, it is easy to check that $x_0:=\left(0,-\overline h\abs{\overline h}^{-3/2}\sqrt{c_0}\right)$ is the unique zero of $f_0$. Replacing it in \eqref{Jacobian}, we see that the determinant is negative, which implies that $0$ is a regular value for $f_0$. Then, by a classical property of the Brouwer degree, we get that
\begin{eqnarray}
d_B(f_0,\Omega\cap\RR^6,0)=\det \text{Jac} f_0 (x_0) = -1 \neq 0.\nonumber
\end{eqnarray}
Finally, applying the global continuation theorem \cite[Theorem 2]{CMZ}, we conclude the proof.

\end{document}